\documentclass[aps,pre,a4,floatfix,groupedaddress,showpacs,showkeys,twocolumn]{revtex4-1}  
\usepackage{graphicx,amsmath,amssymb,dsfont,bm}

\begin{document}

\title{Average cross--responses in correlated financial market}

\author{Shanshan Wang}
\email{shanshan.wang@uni-due.de}
\affiliation{Fakult\"at f\"ur Physik, Universit\"at Duisburg--Essen, Lotharstra\ss e 1, 47048 Duisburg, Germany}
\author{Rudi Sch\"afer}
\email{rudi.schaefer@uni-due.de}
\affiliation{Fakult\"at f\"ur Physik, Universit\"at Duisburg--Essen, Lotharstra\ss e 1, 47048 Duisburg, Germany}
\author{Thomas Guhr}
\email{thomas.guhr@uni-due.de}
\affiliation{Fakult\"at f\"ur Physik, Universit\"at Duisburg--Essen, Lotharstra\ss e 1, 47048 Duisburg, Germany}

\date{\today}

\begin{abstract}
There are non--vanishing price responses across different stocks in correlated financial markets. We further study this issue by performing different averages, which identify active and passive cross--responses. The two average cross--responses show different characteristic dependences on the time lag. The passive cross--response exhibits a shorter response period with sizeable volatilities, while the corresponding period for the active cross--response is longer. The average cross--responses for a given stock are evaluated either with respect to the whole market or to different sectors. Using the response strength, the influences of individual stocks are identified and discussed. Moreover, the various cross--responses as well as the average cross--responses are compared with the self--responses. In contrast, the short memory of trade sign cross--correlation for stock pairs, the sign cross--correlation has long memory when averaged over different pairs of stocks.
\end{abstract}

\maketitle

\section{ Introduction} 
\label{section1}

In recent years, the impact of stock trades on price change has attracted considerable interest~\cite{Chordia2002,Bouchaud2008,Bouchaud2004,Hausman1992,Kempf1999,Dufour2000,Plerou2002,Rosenow2002,Bouchaud2006,Mike2008,Wang2016}. Most of these studies focus on single stocks, because there is a high self--correlation of order flow~\cite{Bouchaud2004,Lillo2004,Lillo2005,Toth2015}. In addition to many stylized facts and specific features~\cite{Chordia2002,Bouchaud2008,Cont2001,Bouchaud2003,Chakraborti2011,Toth2011,Eisler2012,Schmitt2012,Schmitt2013}, the order flow exhibits remarkable persistence. Buy (sell) orders are often followed by more buy (sell) orders, leading to a long--memory self--correlation of trade signs~\cite{Lillo2005}. This is so, because the traders prefer to split large orders into smaller fragments to conceal their trading intentions and to keep the liquidity costs as low as possible. This self--correlation of the order flow generates the stock price self--response to the trades. The price is determined by a continuous double auction~\cite{Farmer2004} with market and limit orders. The market orders are executed immediately at the best available price. Although they do not appear in the order book, they either keep the quote unchanged when the trading volumes are smaller than the market depth or move the price up (down) if they are buyer-- (seller--) initiated. On the other hand the limit orders are executed at the price the traders specified, which is a process that takes time. They are listed in the order book as quotes, \textit{i.e.}  bids and asks, corresponding to the buy and sell limit orders, respectively. The prices change persistently when buy or sell market orders come in, since they remove volumes of the limit orders from the order book and push the price from the best quote to a secondary quote.

Recently, we studied the price cross--response in a correlated market~\cite{Wang2016}. Due to impact mechanisms different from the self--response, the cross--response as well as the cross--correlation of trade signs strongly suffers from noise, leading to drastic fluctuations at large time lags. The sign cross--correlation was thus found to have short memory. For the cross--response, a given stock is related to several or many others. That is partly due to economic dependencies of the companies and to the grouping of investments in portfolios, but there may also be other reasons for the mutual impact. Suppose, for example, a trader who considers a stock as presently underpriced and likely to raise in the near future. To buy many shares of this stock he might use the profit from selling other stocks. If many others act correspondingly, an impact results: buying (selling) this stock affects the other stocks which are sold (bought). By discussing this scenario, we want to motivate that averaging the cross--response functions over different stocks that are paired with the same stock can yield interesting new observations. Furthermore, such averages will also to some extent smoothen the drastic fluctuations of the sign cross--correlations at large time lags and reduce the cross--response noise. We thus introduce the average cross--responses of an individual stock to the whole market and to different economic sectors. In this setting, we also further discuss the influences of individual stocks.

The paper is organized as follows. In Sect.~\ref{section2}, we represent the data sets used for the evaluation of empirical cross--responses and give our definition of the trade signs. In Sect.~\ref{section3}, we introduce the average cross--responses, a passive and an active one, as well as the corresponding cross--correlators of trade signs, and we discuss the influence of zero trade signs. In Sect.~\ref{section4}, we analyze the average cross--responses of an individual stock to the market and to different economic sectors where the two possible definitions, in-- and excluding the zero trade signs, are taken into account. Investigating the average cross--responses, we identify in Sect.~\ref{section5} the influencing and influenced stocks, and analyze the rank difference of influencing stocks with respect to the cross--response. In Sect.~\ref{section6}, we compare the self--response with the various cross--responses. We conclude in Sect.~\ref{section7}.

\vfill

\section{Data description and sign definitions}   
\label{section2}

To make this paper self--contained, we briefly sketch those salient features of Ref.~\cite{Wang2016} which are also needed in the present study. In Sect.~\ref{sec21}, we present the data set that we use in our analysis. In Sect.~\ref{sec22} we give our definition of the trade signs on the physical time scale.

\subsection{Data set}
\label{sec21}

We use the Trades and Quotes (TAQ) data set from NASDAQ stock market of the year 2008. Since the NASDAQ is a purely electronic stock exchange, the information of the stock in each year, such as the trading time, prices and volumes, is conveniently recorded in the TAQ data set. It includes two separate files. One is the trades file with the information of all successive transactions. The other one is the quotes file containing all successive best buy and sell limit orders.

For a stock pair $(i,j)$, only the common trading days are taken into account, because an intraday price cross--response does not occur if one of the stocks is not traded on a given day. Moreover, to avoid artifacts related to the dramatic price fluctuations at the opening and closing of the market and to overnight effects, the trading time we consider for all stocks is from 9:40:00 to 15:50:00 New York local time.

For the average cross--responses of an individual stock in Sect.~\ref{section4}, Apple Inc. (AAPL), Goldman Sachs Group (GS), and Exxon Mobil Corp. (XOM) are selected as sample stocks, where AAPL belongs to the economic sector of Information technology, GS belongs to Financials, and XOM to Energy. The averages are performed over the other available 495 stocks in the S$\&$P 500 index or over the stocks in a given economic sector. We always omit the self--response of the stocks in the averages.

When analyzing the influencing and influenced stocks using the average cross--responses in Sect.~\ref{section5}, in total 99 stocks from the S$\&$P 500 index in 2008 are selected, as listed in App.~\ref{appA}. The 99 stocks include the first ten stocks with the largest average market capitalization in each economic sector, except for the telecommunications services with only nine available stocks in that year. We recall that the average market capitalization is obtained as the traded price multiplied with the traded volume, averaged over every trade during the year 2008. In addition, the 99 stocks are also used to work out the $99\times99$ cross--response matrix in Sect.~\ref{sec33}.

\subsection{Definition of trade signs}
\label{sec22}

As each stock has its own trading time, which never is synchronous with that of another stock, it is not useful to work out the cross--response between stocks on a trading time scale. Instead, a discrete physical time scale is appropriate, taking into account the trades at the actual time stamp. Thereby, all cross--responses are treated on equal footing. The time stamp in our study is set to be one second which is the minimal time scale given in the TAQ data set. Since more than one trade or quote may be recorded in the same second, the trade sign at one second is an aggregated sign of all the trades in this second. If the consecutive time intervals of one second are labeled by $t$ and the total number of trades in time $t$ is denoted by $N(t)$, the trade sign for one second is defined as
\begin{eqnarray}       
\varepsilon(t)=\left\{                  
\begin{array}{lll}    
\mathrm{sgn}
\left(\sum\limits_{n=1}^{N(t)}\varepsilon(t;n)\right) & \ , \quad & \mbox{if} \quad N(t)>0 \ , \\    
                                                0 & \ , \quad & \mbox{if} \quad N(t)= 0 \ .
\end{array}           
\right. 
\label{eq21}              
\end{eqnarray}     
Here, $\varepsilon(t;n)$ is the trade sign of the $n$--th trade in the time interval $t$ of length one second and the function $\textrm{sgn}$ returns the sign of the argument. A non--zero value of $\varepsilon(t)$ indicates the number imbalance of trades in time $t$. To be specific, $\varepsilon(t)=+1$ implies that there was a majority of buy market orders in time $t$, while $\varepsilon(t)=-1$ indicates a majority of sell market orders. The value $\varepsilon(t)=0$ either means that there was not any trade executed in time $t$ or that there was a balance of buy and sell market orders.

To define the sign of each trade, \textit{i.e.} $\varepsilon(t;n)$, there are three possible approaches. The most popular one was proposed by Lee and Ready~\cite{Lee1991}. The trade sign is determined by comparing the traded price with the preceding midpoint price of the best buy and sell quotes. If the traded price is larger than the preceding midpoint price, the trade is buyer--initiated, and otherwise seller--initiated. The Lee and Ready algorithm is able to correctly classify at least 85$\%$ of all the trades. Unfortunately, as the trades and quotes are recorded in two separate files on time scale of one second, \textit{i.e.}, not distinguishing what happens within these one--second intervals, it is not always possible to associate the trades with their preceding quotes at time scales smaller than one second. Another approach is put forward by Holthausen, Leftwich and Mayers~\cite{Holthausen1987}. The trade is classified as buyer-- or seller--initiated if the traded price is different from the previous price. The advantage is that the trading sequence can be identified only using the trades file. However, as the trade is not classified when the two consecutive prices are the same, this approach has a rather low accuracy of 52.8$\%$. Recently, we put forward a third approach~\cite{Wang2016} in which a trade with the same price as the preceding one is also classified. Compared to the second approach, it has a considerably higher accuracy of 85$\%$. This number is based on tests carried out with six samples, where in total 308857 transactions were identified. Here, we again employ the third approach, in which the trade sign is defined as
\begin{eqnarray}       
\varepsilon(t;n)=\left\{                  
\begin{array}{lll}    
\mathrm{sgn}\bigl(S(t;n)&-&S(t;n-1)\bigr)   \ ,  \\
                                      &&\mbox{if}~~S(t;n)\neq S(t;n-1) , \\    
            \varepsilon(t;n-1) \ ,&&\mbox{otherwise},
\end{array}           
\right.  
\label{eq22}            
\end{eqnarray}
where, $S(t;n)$ denotes the price of the $n$--th trade in the one--second time interval $t$. The traded price is the same as the preceding one if there was sufficient volume at the best quote that was not exhausted by two or more consecutive buy or sell trades within $t$.

\section{Average cross--response functions}
\label{section3}

We introduce the passive and active cross--response functions in Sect.~\ref{sec31} and correspondingly the passive and active sign cross--correlators in Sect.~\ref{sec32}. We then discuss the influence of zero trade signs on the cross--response of individual stocks in the financial market in Sect.~\ref{sec33}.

\subsection{Passive and active cross--response functions} 
\label{sec31}

The cross--response function for a stock pair $(i,j)$ is defined as~\cite{Wang2016}
\begin{equation}
R_{ij}(\tau) \ = \ \Bigl\langle r_i(t,\tau)\varepsilon_j(t)\Bigr\rangle _t \ , 
\label{eq31}
\end{equation}
Here, $\langle\cdots\rangle_t$ represents the average with respect to time $t$. Moreover, $r_i(t,\tau)$ denotes the logarithmic price change, defined by the midpoint price $m_i(t)$,
\begin{equation}
r_i(t,\tau) \ = \ \log m_i(t+\tau) -\log m_i(t) \ = \ \log\frac{m_i(t+\tau)}{m_i(t)} \ .
\label{eq32}
\end{equation}
Thus, the cross--response function for a stock pair measures how the buy or sell market orders of stock $j$ at time $t$ impact the price of stock $i$ after a time lag $\tau$. In correlated financial market, the price of stock $i$ might be impacted by multiple stocks simultaneously. Likewise, the trades of stock $j$ might also impact on multiple stocks concurrently. As the definition Eq.~\eqref{eq31} of the cross--response function is not symmetric with respect to the indices, we can perform two conceptually different averages,
\begin{equation}
R_i^{(p)}(\tau)=\left\langle R_{ij}(\tau)\right\rangle_{j}
\quad \textrm{and} \quad
R_j^{(a)}(\tau)=\left\langle R_{ij}(\tau)\right\rangle_{i},
\label{eq33}
\end{equation}
to which we refer as passive and active cross--response functions. Importantly, the self--responses for $(i,i)$ or $(j,j)$ are excluded in these averages.  The passive cross--response function $R_i^{(p)}(\tau)$ measures, how the price of stock $i$ changes due to the trades of all other stocks, while the active cross--response function $R_j^{(a)}(\tau)$ quantifies which effect the trades of stock $j$ has on the prices of all other stocks.

\subsection{Passive and active sign cross--correlators}
\label{sec32}

The trade sign cross--correlator for the stock pair $(i,j)$ is given by~\cite{Wang2016},
\begin{equation}
\Theta_{ij}(\tau) \ = \ \Bigl\langle \varepsilon_i(t+\tau)\varepsilon_j(t) \Bigr\rangle _t \ .
\label{eq34}
\end{equation}
The empirical results of the sign cross--correlation can be fitted well by a power law function,
\begin{equation}
\Theta_{ij}(\tau)=\frac{\vartheta_{ij}}{\left(1+(\tau/\tau_{ij}^{(0)})^2\right)^{\gamma_{ij}/2}} \ ,
\label{eq35}
\end{equation}
with constants $\vartheta_{ij}$, $\tau_{ij}^{(0)}$ and $\gamma_{ij}$. We found a short--memory with exponent $\gamma_{ij}\geq 1$ for the self--correlator as well as for the cross--correlators of trade signs on the physical time scale~\cite{Wang2016}. However, previous studies~\cite{Bouchaud2004,Lillo2004} yielded a long--memory self--correlator with $\gamma<1$ on the trading time scale. The difference for the self--correlator is due to the different time scales used for the measurements. In the case of the cross--correlator, there are strong fluctuations at large time lags. They mainly result from the trading noise for a stock pair. To reduce this noise, we average the cross--correlator of trade signs over different stock pairs. Thus, corresponding to the passive and active cross--responses, we introduce
\begin{equation}
\Theta_i^{(p)}(\tau) = \left\langle \Theta_{ij}(\tau)\right\rangle_{j}
\quad \textrm{and} \quad
\Theta_j^{(a)}(\tau) = \left\langle \Theta_{ij}(\tau)\right\rangle_{i} ,
\label{eq36}
\end{equation}
as passive and active trade sign cross--correlators. We notice that they are not symmetric either, because the time lag only enters the trade sign with index $i$. The memory properties for these two types of average sign cross--correlations are also checked with a power law function of the type~\eqref{eq35} in Sect.~\ref{section4}.

\subsection{Influence of zero trade signs}
\label{sec33}

\begin{figure}[tbp]
  \begin{center}
    \includegraphics[width=0.48\textwidth]{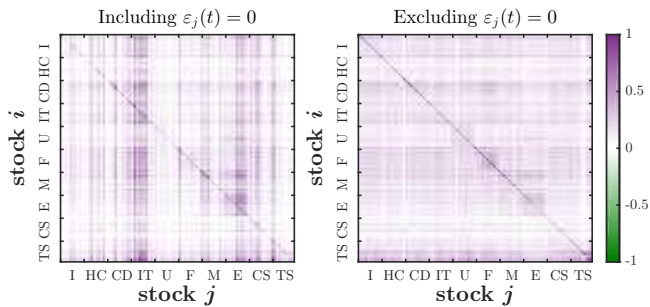}
  \end{center}
  \vspace*{-0.3cm}
  \caption{Matrices of the market response with entries $\rho_{ij}(\tau)$ for $i,j=1,\ldots,99$ at the time lag $\tau=60$~s in 2008. The stocks pairs $(i,j)$ belong to the sectors industrials (I), health care (HC), consumer discretionary (CD), information technology (IT), utilities (U), financials (F), materials (M), energy (E), consumer staples (CS), and telecommunications services (TS).}
  \vspace*{-0.3cm}
 \label{fig31}
\end{figure}

The lack of trading and the balance of buy and sell market orders within one second leads to zero trade signs $\varepsilon_j(t)=0$ on the physical time scale. In Ref.~\cite{Wang2016}, we studied the effects due to in-- or excluding these events. The cross--responses including $\varepsilon_j(t)=0$ are weaker than the ones excluding $\varepsilon_j(t)=0$, because the latter is purely built up by trades, and the indirect influence of the lack of trades is omitted. However, either including or excluding $\varepsilon_j(t)=0$ does not change the trend of price reversion versus the time lag, but it does affect the cross--response strength. As a result, the market displays, at a given time lag $\tau$, distinct cross--response structures. To analyze them, we defined~\cite{Wang2016} the normalized cross--responses,
\begin{equation}
\rho_{ij}(\tau) \ = \ \frac{R_{ij}(\tau)}{\textrm{max\,}(|R_{ij}(\tau)|)} \ ,
\label{eq37}
\end{equation}
where the denominator is the maximum over all stock pairs $(i,j)$ for fixed $\tau$. The resulting matrices in-- or excluding $\varepsilon_j(t)=0$ are shown in Fig.~\ref{fig31}.  When including $\varepsilon_j(t)=0$, some stocks acting as the impacting stocks $j$ dominate the market, such as the stocks from information technology (IT) sector. This yields the strip feature in the market response structure. When excluding $\varepsilon_j(t)=0$, we do not find such a striking impacting stock $j$. Moreover, the changes in the cross--responses across the impacted stocks $i$ are comparable to that across the impacting stocks $j$.  The distinct market cross--response structures provide different market information dependent of whether to take into account the influence of $\varepsilon_j(t)=0$ or not. As this may also produce different average cross--responses of an individual stock across the financial market, we keep both cases in our study.

\section{Average cross--responses of an individual stock}
\label{section4} 

We analyze the average cross--responses of the individual stocks to the whole market in Sect.~\ref{sec41} and to different economic sectors in Sect.~\ref{sec42}. 

\subsection{Average cross--responses to the market}
\label{sec41}

\begin{figure*}[htbp]
  \begin{center}
    \includegraphics[width=0.8\textwidth]{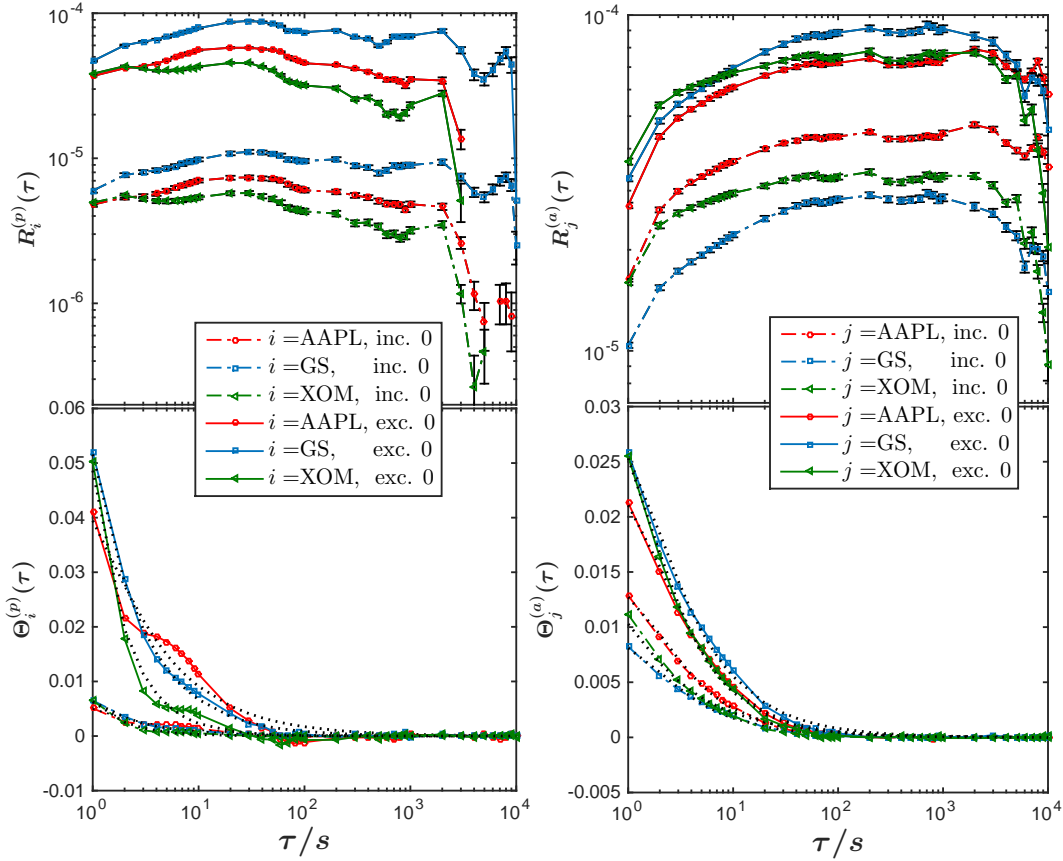}
    \vspace{-0.5cm}
        \caption{Passive and active cross--response functions $R_i^{(p)}(\tau)$ and $R_j^{(a)}(\tau)$ for $i,j=\textrm{AAPL},\textrm{GS},\textrm{XOM}$ in the year 2008 versus time lag $\tau$ on a logarithmic scale (top, left and right). Corresponding passive and active trade sign cross--correlators $\Theta_i^{(p)}(\tau)$ and $\Theta_j^{(a)}(\tau)$, fit as dotted lines (bottom, left and right). The error bars indicate the standard errors.}
   \label{fig41}
  \end{center}
\end{figure*}

\begin{table*}[htbp]
\linespread{0.5} 
  \caption{Fit parameters and errors $\chi_{i}^2$ or $\chi_{j}^2$~\cite{Wang2016} for the average trade sign cross--correlators.} 
\begin{center}
\begin{tabular}{c@{\hskip 0.2in}l@{\hskip 0.2in}c@{\hskip 0.1in}cc@{\hskip 0.2in}c@{\hskip 0.1in}cc@{\hskip 0.2in}c@{\hskip 0.1in}cc@{\hskip 0.2in}cc} 
\hline
\hline
Sign &stock $i$, $j$ 	& \multicolumn{2}{c}{$\vartheta_{i}$ or $\vartheta_{j}$} && \multicolumn{2}{c}{$\tau_{i}^{(0)}$ or $\tau_{j}^{(0)}$~[ s ]}&&\multicolumn{2}{c}{$\gamma_{i}$ or $\gamma_{j}$ }& & \multicolumn{2}{c}{$\chi_{i}^2$ or $\chi_{j}^2$~($\times10^{-6}$)} \\
\cline{3-4}\cline{6-7}\cline{9-10}\cline{12-13}
cross--correlators		&			&inc. 0	&exc. 0	&&inc. 0	&exc. 0	&&inc. 0	&exc. 0	&&inc. 0	&exc. 0	\\
\hline
					&AAPL 		&0.01	&0.05	&&0.47	&0.88	&&0.68 	&0.73	&&0.07 	&4.59	\\
$\Theta_i^{(p)}(\tau)$	&GS			&0.03	&0.22	&&0.23	&0.20	&&0.92 	&0.90	&&0.01	&0.38 	\\
					&XOM		&0.27	&0.83	&&0.06	&0.12	&&1.32	&1.33	&&0.02	&1.20	\\
\hline					
					&AAPL		&0.02	&0.03	&&1.44	&1.44	&&0.90 	&0.91	&&0.03 	&0.08	\\
$\Theta_j^{(a)}(\tau)$	&GS			&0.01	&0.03	&&1.31 	&1.27	&&0.85 	&0.83	&&0.02 	&0.18	\\
					&XOM		&0.02	&0.03	&&0.55	&1.08	&&0.71	&0.95	&&0.11	&0.08	\\
\hline
\hline
\end{tabular}
\end{center}
\label{tab41}
\end{table*}

We carry out the empirical analysis for the stocks AAPL, GS, XOM by averaging their cross--response functions over other 495 stocks in the S$\&$P500 index. The results for passive and active cross--response functions as well as the corresponding passive and active trade sign cross--correlators are presented in Fig.~\ref{fig41}. We distinguish the results for in-- or excluding $\varepsilon_j(t)=0$. We checked that these empirical results are similar to those from averaging over the in each case remaining 98 stocks listed in App.~\ref{appA}. To facilitate the computation, we only calculate the average cross--response values at several time lags, which are marked in Fig.~\ref{fig41}, rather than at every second.

The passive and active cross--response functions clearly show different behaviors. The passive cross--response reverses faster than the active one. It only persists dozens of seconds and then reverses to drop down quickly with sizeable volatility. In contrast, the active cross--response reverses at time lags of some hundreds of seconds and the price changes slowly. An obvious reason for this difference is the easier detectability of price changes in one stock than of those dispersed over different stocks. Again, with our results, we can extend the previous interpretations based on the study of single stocks.  The passive cross--response function reflects the price dynamics on short time scales. When the price goes up, less market orders to buy will be emitted and more limit orders to sell. Thus, the price reverses~\cite{Hopman2007} without a need to evoke new information as a cause. Moreover, liquidity induced mean reversion attracts more buyers, which motivates liquidity providers to raise the price again, while the volatilities in this process of responding decline. Thus, we conclude for the market as a whole that the mean reversion accentuates the short--period price volatility, which is consistent with the single--stock analysis~\cite{Handa1998,Bouchaud2010}. The active cross--response reflects the dispersion of the trade impacts over the prices of different stocks. It is conceivable that this process takes longer compared to the time scales of the passive cross--response. Furthermore, this dispersion is accompanied by spreading out the volatilities.

As visible in Fig.~\ref{fig41}, for the average cross--responses including $\varepsilon_j(t)=0$, the active ones are all stronger than the passive ones at their maximum values. The reason for the different strengths of passive and active cross--responses is the existence of strongly influencing stocks. On the left hand side of Fig.~\ref{fig31}, we observe that the vertical stripes are much more pronounced than the horizontal ones. More specifically, there are groups of stocks which have a strong influence across most of the market, in particular, this is evident for the stocks in the IT sector. Consequently, the active cross--response of these stocks averaged over the market shows a strong signal. As for the passive cross--response, however, fewer influenced stocks contribute, which leads to a reduced value of the average.  However, the strength difference is not so apparent for the average cross--responses excluding $\varepsilon_j(t)=0$. The right hand side of Fig.~\ref{fig31} shows that the influencing and influenced stocks contribute to the average cross--responses with similar strength.

To analyze the average trade sign cross--correlators depicted in Fig.~\ref{fig41}, we use the power law~\eqref{eq35} to fit the empirical results. The fitting parameters and errors are shown in Table~\ref{tab41}. The remarkable result is that the volatile short memory of the individual cross--correlators~\cite{Wang2016} turns into a long memory with exponents smaller than one after averaging. The only exception is the passive trade sign cross--correlator for the stock XOM. We thus infer that the price changes caused by trade sign cross--correlations in different stocks can accumulate to persist over longer time. Another interesting aspect is the decay period $\tau_i^{(0)}$ or $\tau_j^{(0)}$ of the sign cross--correlators. For stock pairs, the decay period excluding $\varepsilon_j(t)=0$ is longer than the one including $\varepsilon_j(t)=0$~\cite{Wang2016}. This might be attributed to the influence of the lack of trading or to the balance of buy and sell market orders, which both accelerate the decay of sign cross--correlations. By averaging the cross--response across different stocks, however, the two decay periods are effectively enhanced and become comparable. This is so because the noise is reduced by averaging, slowing down the decay and yielding long--memory sign cross--correlations.

\subsection{Average cross--responses to economic sectors}
\label{sec42}

\begin{figure*}[htbp]
  \begin{center}
    \includegraphics[width=0.95\textwidth]{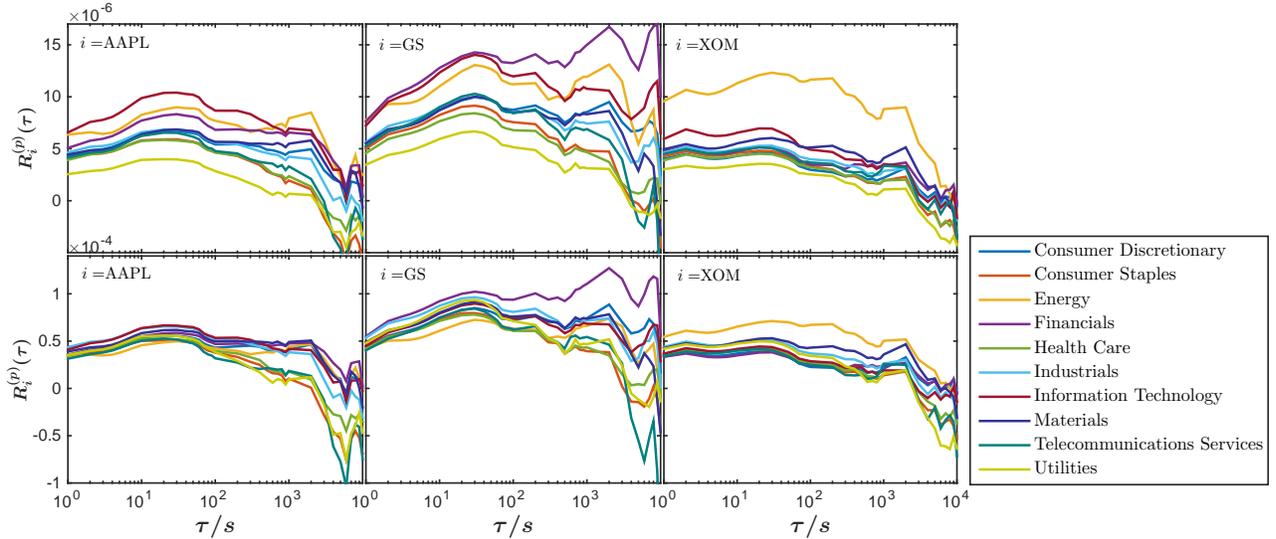}
  \end{center}
   \vspace*{-0.3cm}
   \caption{Passive cross--response functions $R_i^{(p)}(\tau)$ of the stocks $i=\textrm{AAPL},\textrm{GS},\textrm{XOM}$ to ten different economic sectors in the year 2008 versus time lag $\tau$ on a logarithmic scale: in-- and excluding $\varepsilon_j(t)=0$ in the top and bottom panels, respectively.}
  \label{fig42} 
\end{figure*}

\begin{figure*}[htbp]
  \begin{center}
    \includegraphics[width=0.95\textwidth]{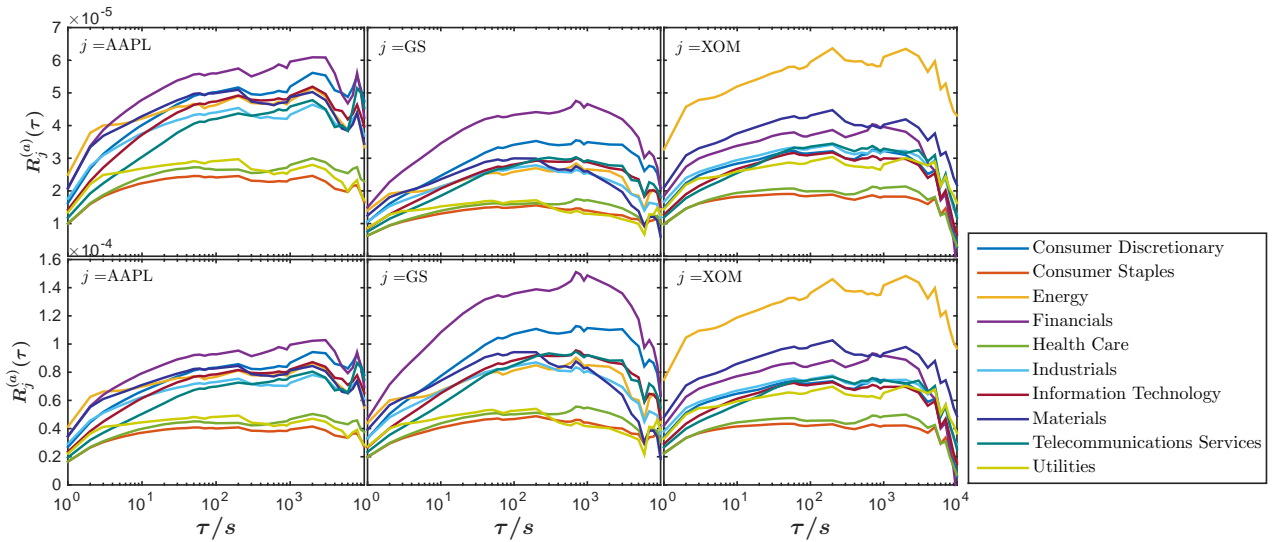}
  \end{center}
   \vspace*{-0.3cm}
   \caption{Active cross--response functions $R_j^{(a)}(\tau)$ of the stocks $j=\textrm{AAPL},\textrm{GS},\textrm{XOM}$ to ten different economic sectors in the year 2008 versus time lag $\tau$ on a logarithmic scale: in-- and excluding $\varepsilon_j(t)=0$ in the top and bottom panels, respectively.}
 \label{fig43}
\end{figure*}

Another observation which can be made in Fig.~\ref{fig31} is that the cross--responses vary for different economic sectors. In other words, the stocks from different sectors may produce different average cross--responses to a given stock. We calculate the average cross--responses of the stocks AAPL, GS and XOM to ten economic sectors in the S$\&$P 500 index. The passive and active cross--responses are displayed in Figs.~\ref{fig42} and~\ref{fig43}, respectively. Both of the average cross--responses show a strong dependence on the economic sectors. Here, we also consider the cross--responses in-- and and excluding $\varepsilon_j(t)=0$, and we notice a remarkable difference for the passive cross--responses.  The three stocks considered have in common that their prices are all affected by the trades within their own sectors, especially XOM, which is not surprising due to the same economic effects.

The active cross--responses, for $\varepsilon_j(t)=0$ in-- and excluded, show time--dependent clustering across different sectors. This will be further discussed in Sect.~\ref{sec53}. The trades of AAPL and GS have a significant impact on the prices of the stocks from financials (F), but a lesser one on utilities (U), health care (HC), and consumer staples (CS). This might be due to the stability of these sectors which serve the needs of daily life. The trades of XOM are more likely to influence energy (E), but have a lesser effect on health care (HC) and consumer staples (CS). This is so because utilities (U) are economically more strongly coupled to energy (E) than to health care (HC) and consumer staples (CS).

\section{Influencing and influenced stock from the viewpoint of average cross--responses}
\label{section5}

\begin{figure}[t]
  \begin{center}
    \includegraphics[width=0.49\textwidth]{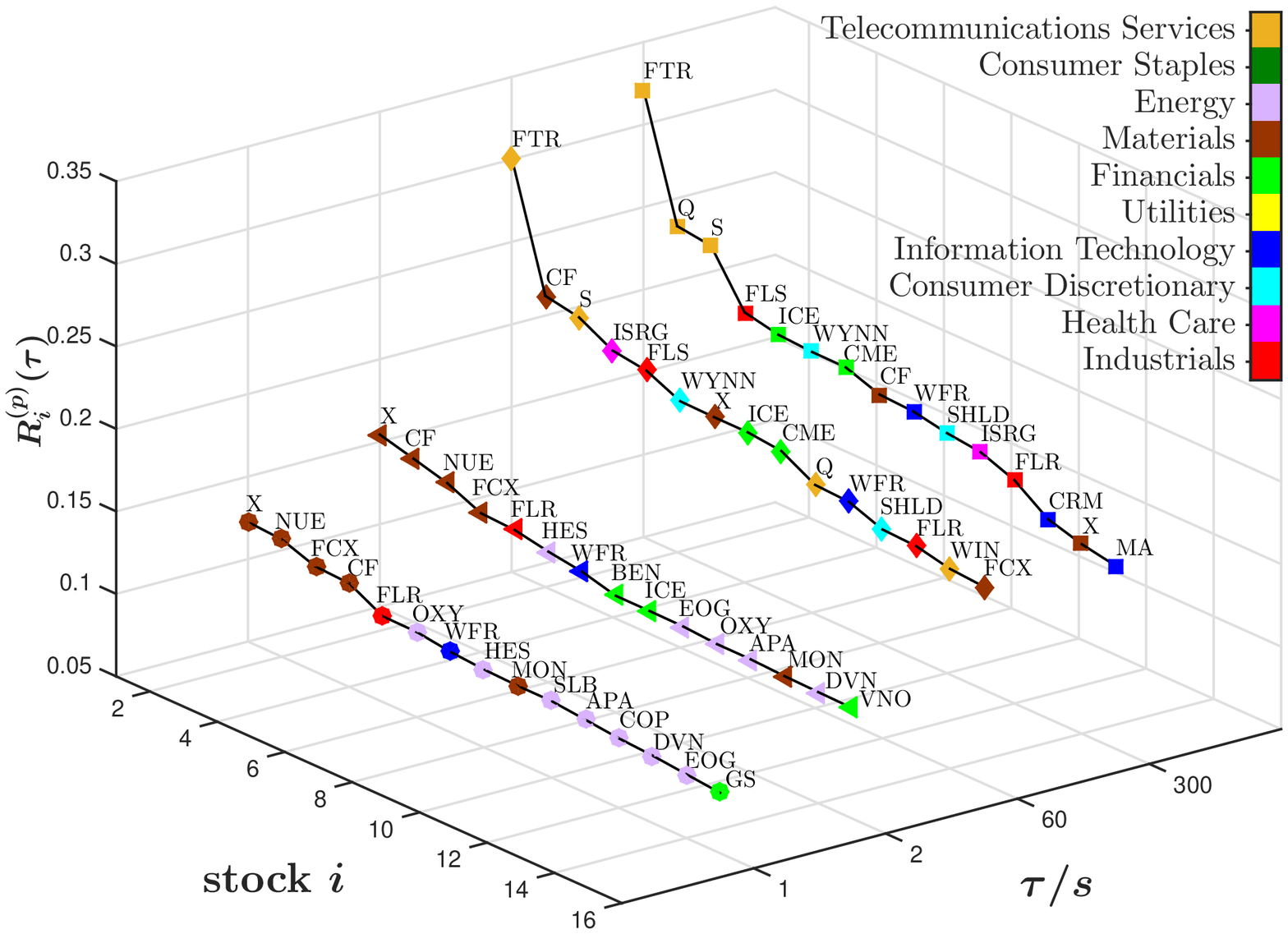}
    \includegraphics[width=0.49\textwidth]{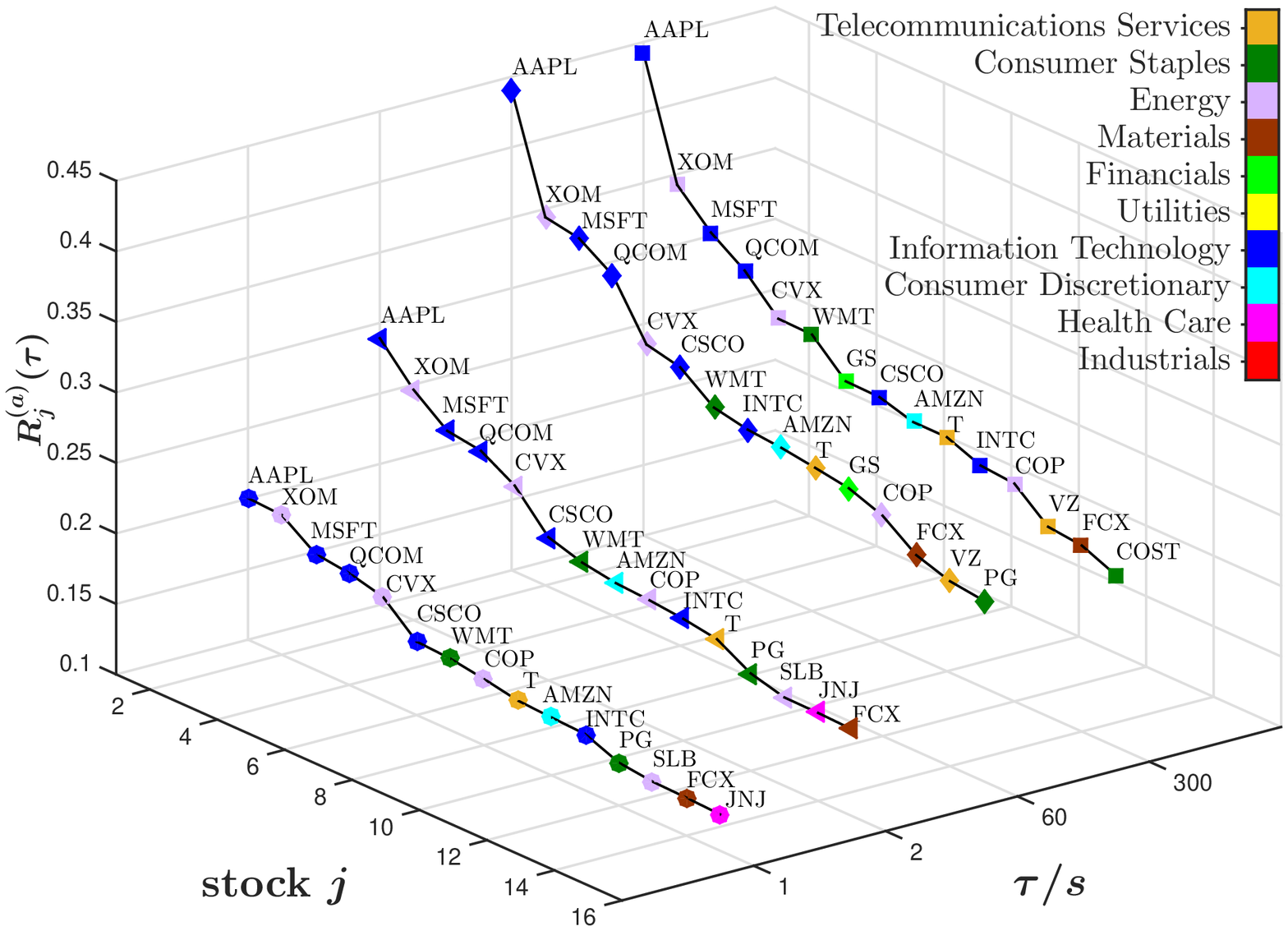}
  \end{center}
   \vspace*{-0.2cm}
   \caption{The first fifteen stocks with the strongest passive (top) and active (bottom) cross--response functions $R_i^{(p)}(\tau)$ and $R_j^{(a)}(\tau)$ versus stock index $i$ or $j$ and time lags $\tau=1$~s ($\circ$), 2~s ($\triangleleft$), 60~s ($\diamond$), and 300~s ({$\Box$}). The cross--response functions include $\varepsilon_j(t)=0$. The ordinates of top and bottom graphs extend over different intervals.}
 \label{fig51}
\end{figure}

\begin{figure}[t]
  \begin{center}
    \includegraphics[width=0.49\textwidth]{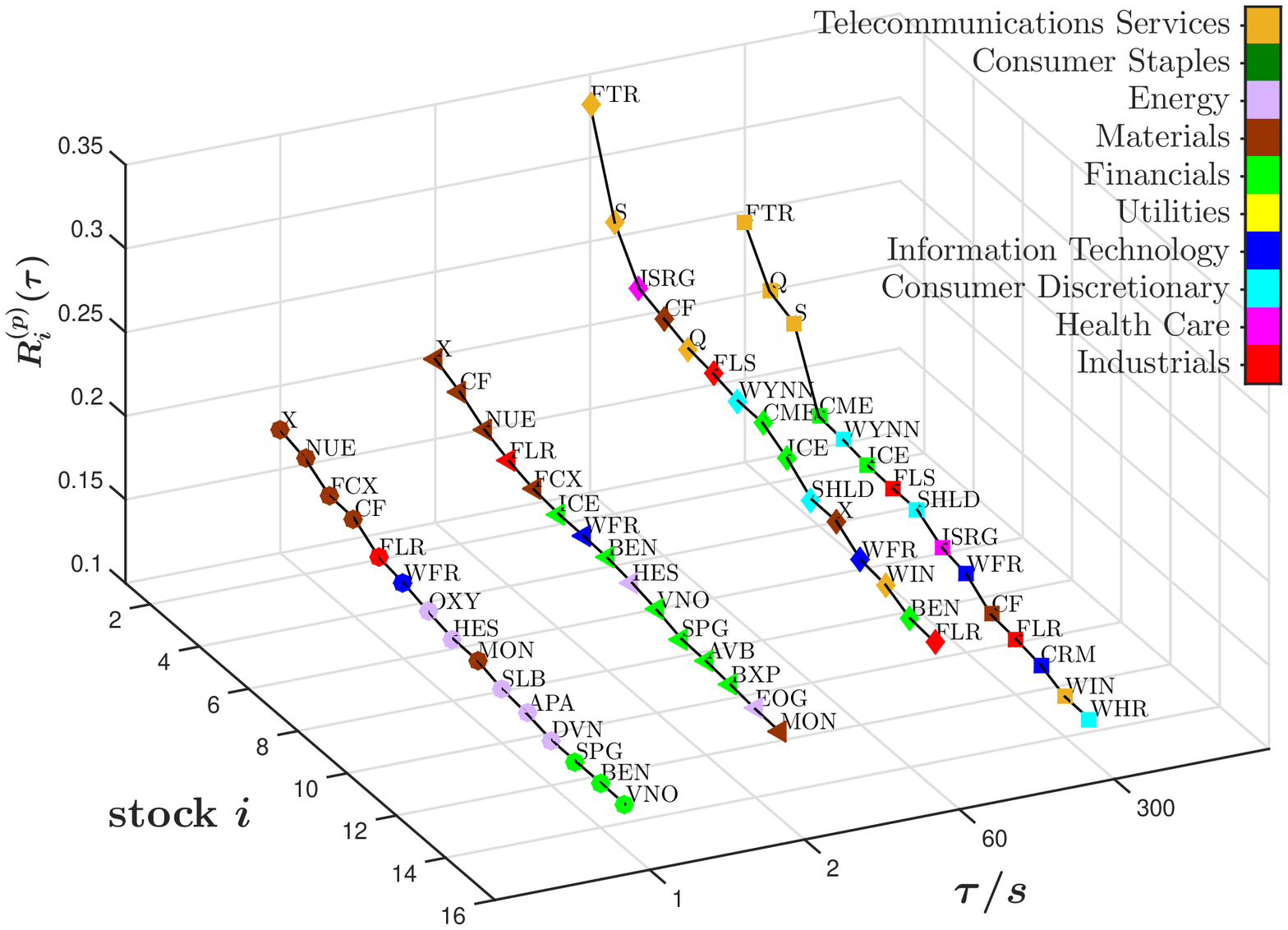}
    \includegraphics[width=0.49\textwidth]{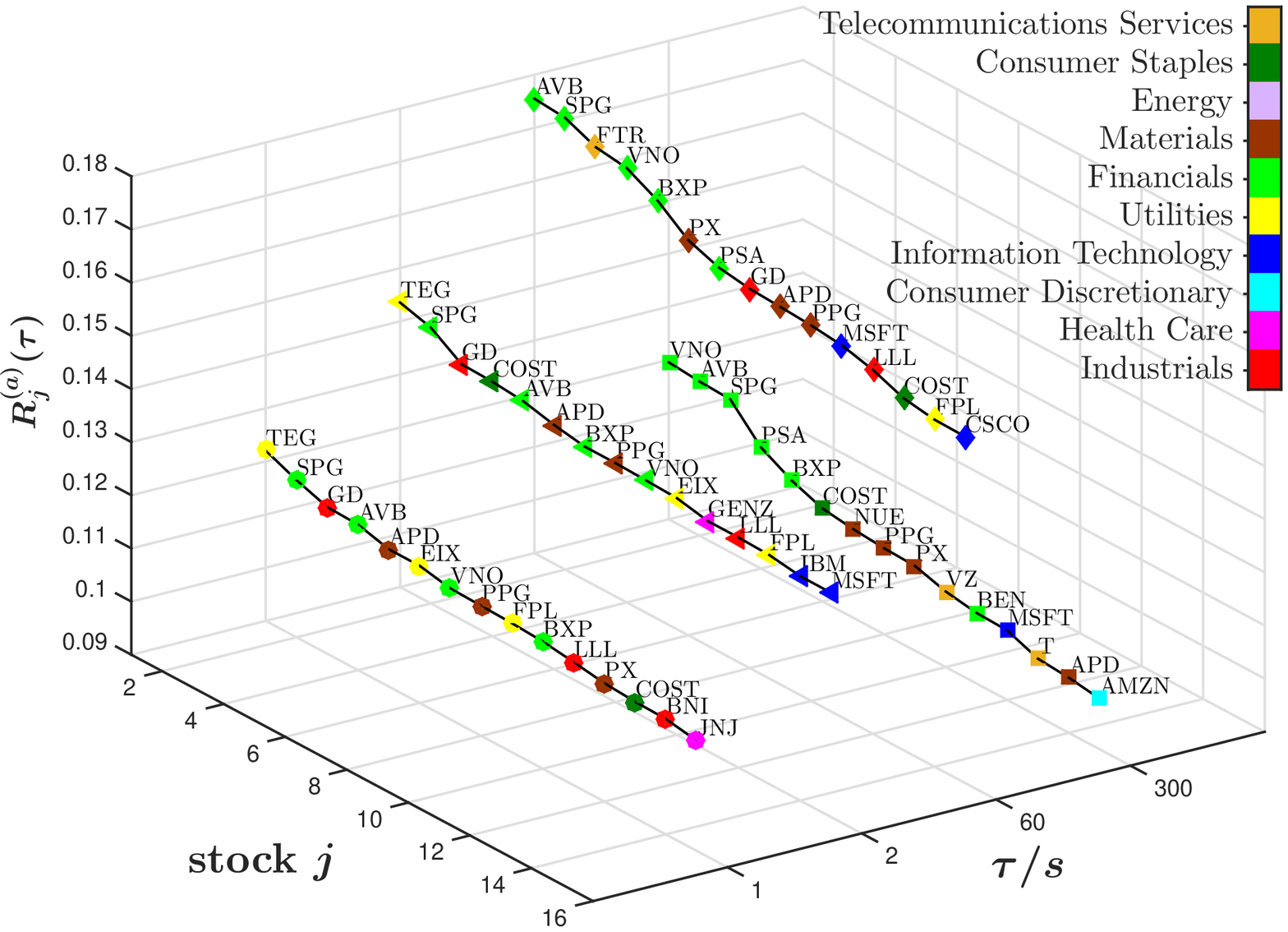}
  \end{center}
   \vspace*{-0.2cm}
   \caption{The first fifteen stocks with the strongest passive (top) and active (bottom) cross--response functions $R_i^{(p)}(\tau)$ and $R_j^{(a)}(\tau)$ versus stock index $i$ or $j$ and time lags $\tau=1$~s ($\circ$), 2~s ($\triangleleft$), 60~s ($\diamond$), and 300~s ({$\Box$}). The cross--response functions exclude $\varepsilon_j(t)=0$. The ordinates of top and bottom graphs extend over different intervals.}
 \label{fig52}
\end{figure}

We identify the influencing and influenced stocks respectively employing active and passive cross--responses in Sect. \ref{sec51}.  We discuss the relations between influencing stocks and trading frequency in Sect.~\ref{sec52}, and we analyze the relation of average cross--responses to the trading frequency in Sect.~\ref{sec53}.

\subsection{Identifying the influencing and influenced stocks}
\label{sec51}

\begin{figure*}[htbp]
  \begin{center}
    \includegraphics[width=0.45\textwidth]{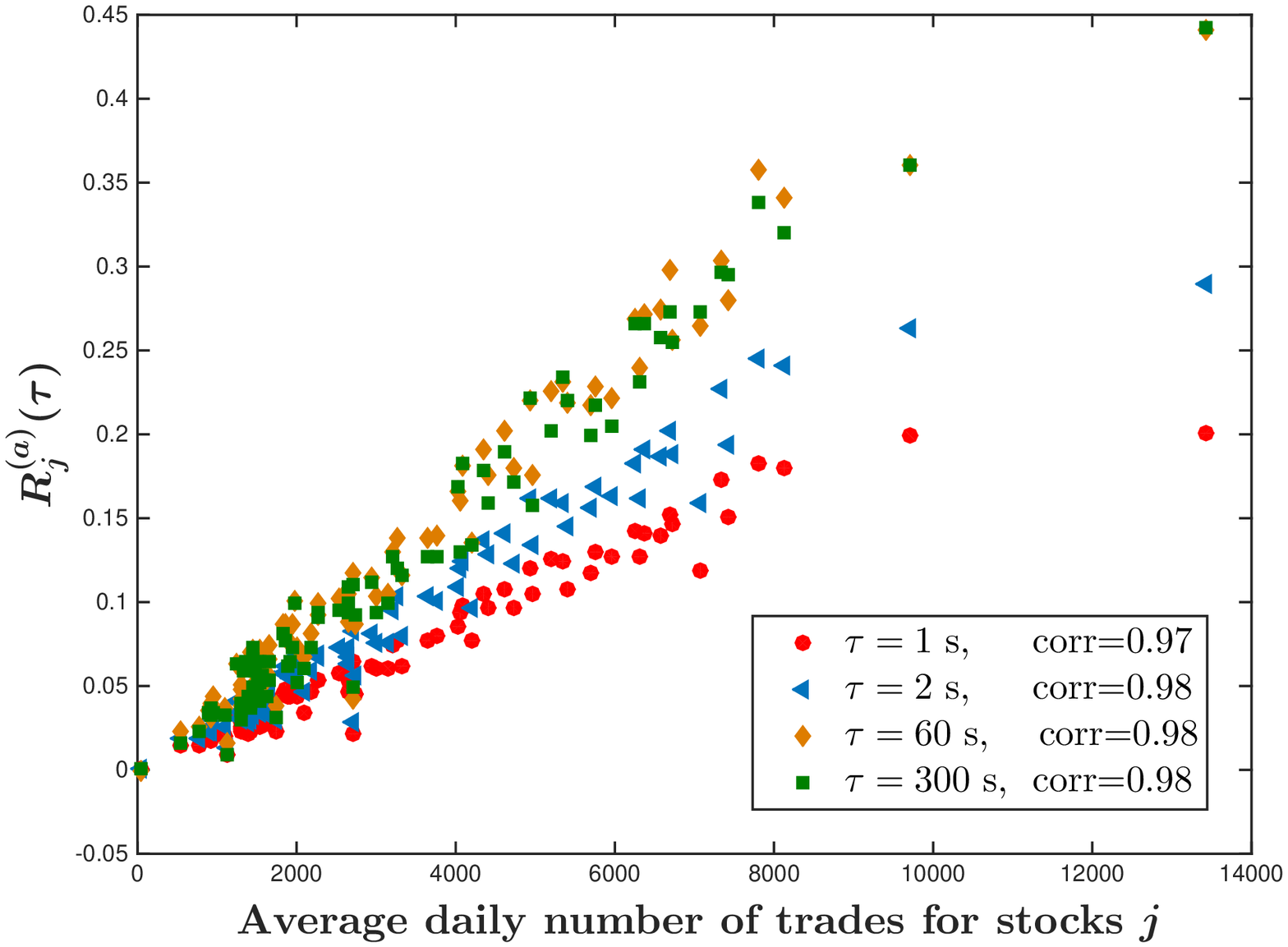}
     \includegraphics[width=0.45\textwidth]{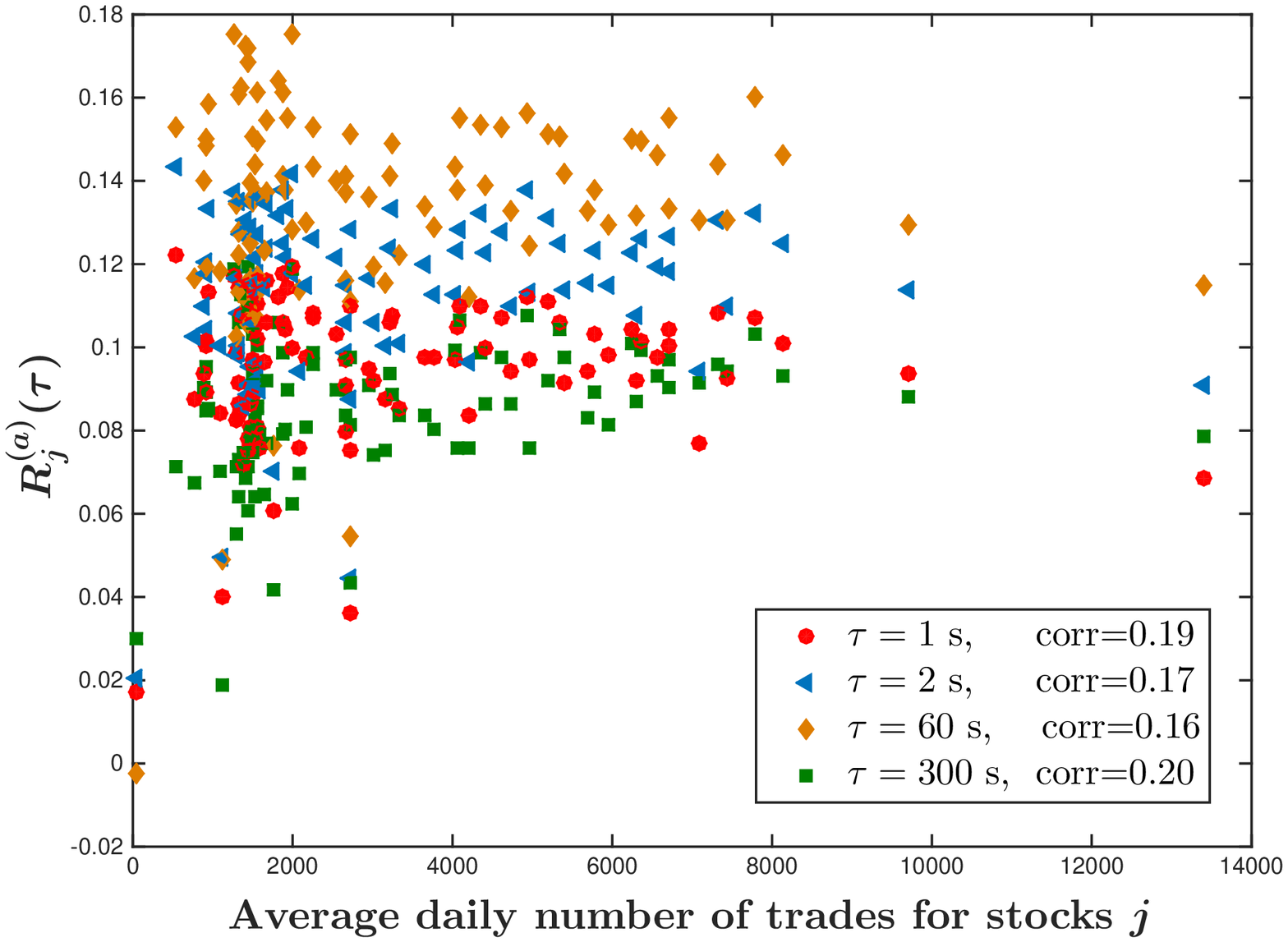}
  \end{center}
   \vspace*{-0.5cm}
   \caption{Relation between active cross--response and average daily number of trades for the stocks $j$. The Pearson correlation coefficient of these two quantities is denoted by corr.  The number of daily trades is at most one per second from 9:40 to 15:50 New York local time. The cross--response functions in-- and exclude $\varepsilon_j(t)=0$ in the left and the right panels, respectively.}
 \label{fig53}
\end{figure*}

As the passive and active cross--response functions measure the strength with which one stock impacts or is impacted by other stocks, respectively, we can classify the stocks as the influencing and influenced stocks. The influenced stocks with strong passive cross--responses are easily impacted by the trades of other stocks. On the other hand, the influencing stocks with strong active cross--responses have large impacts on the prices of other stocks. Again, the impact here is separated into the impacts in-- or excluding zero trade signs.

To identify the strongly influencing and influenced sto\-cks, we rank the 99 stocks listed in App.~\ref{appA} according to the numerical values of passive and active cross--response functions, normalized according to Eq.~\eqref{eq37}, at a given time lag $\tau$.  The first fifteen stocks with strongest average cross--responses at $\tau=1, 2, 60, 300$~s are shown in Fig.~\ref{fig51} including $\varepsilon_j(t)=0$ and in Fig.~\ref{fig52} excluding $\varepsilon_j(t)=0$. As seen, for the passive cross--response either in-- or excluding $\varepsilon_j(t)=0$, the rank of strongly influenced stocks is not changed too much, especially for FTR, X, CF, NUE and S. This is so, because the influence of zero trade signs is reduced across different stocks, altering the ranks of influenced stocks only little. However, the influence is spread to different stocks simply due to trading. Thus great difference of the ranks for the influencing stocks can be found in the active cross--responses in-- and excluding $\varepsilon_j(t)=0$. As for the former, most of the influencing stocks come from the information technology (IT) sector. Their ranks look similar at different time lags. This matches the relatively stable cross--response structure visible in Fig.~\ref{fig31}. As for the latter, more stocks from financials (F) are in the top positions. Even some stocks from utilities, which are not in the top ranks in the former case, are identified as the influencing stocks.

\subsection{Relations between influencing stocks and trading frequency}
\label{sec52}

The zero trade signs relate to the trading frequency of the stocks on the time scale of one second. The fewer zero trade signs, the higher is the trading frequency. Hence, the ranks of influencing and influenced stocks are connected to the trading frequency of those stocks. Because of the great difference in the ranks for influencing stocks when comparing the in-- and exclusion of zero trade signs, we check the relation between the active cross--responses of individual stocks and their average daily number of trades. The number of daily trades is at most one per second from 9:40 to 15:50 New York local time. It is the sum of the absolute values of the trade signs defined in Eq.~\eqref{eq21}. The 99 stocks listed in App.~\ref{appA} are used for the analysis.

In Fig.~\ref{fig53}, two different scenarios are visible. For $\varepsilon_j(t)=0$ included, there is a linear relation with a high correlation between the active cross--response and the average daily number of trades. Thus, the influencing stocks are also the stocks with high trading frequency. The impacts of trades from the frequently traded stocks rapidly follow each other. Before the previous impact vanishes, the new one appears and pushes the price persistently. The cross--responses of the infrequently traded stocks are weaker, because it happens that the previous impact has vanished, but the next impact has not arrived yet. As a result, the prices of other stocks are affected more by other information, \textit{e.g.} news, rather than by the impact of trades. On the other hand, when $\varepsilon_j(t)=0$ is excluded, the linear relation disappears. The high trading frequency does not longer indicate the high influence of individual stocks. The strongly influencing stocks are more likely to be found in the stocks with an average daily number of trades smaller than 2000.

\subsection{Relations of average cross--responses in-- and excluding zero trade signs}
\label{sec53}

As we know, the asynchronous trading induces spurious lead--lag correlations, and the returns of frequently traded stocks generally lead those of infrequently traded stocks \cite{Lo1990a,Lo1990b}. For example, incoming news influences the frequently traded stocks first, and then, after a time lag, the infrequently traded stocks. In our study, the high correlation between the active cross--response including $\varepsilon_j(t)=0$ and the average daily number of trades might be misinterpreted such that the influence that the stocks exert is due to the lead--lag correlation of the return rather than due to the impact of trades. When including the zero trade signs, we take into account how the lack of trading or the balance of buy and sell market orders affects the impact of trades. Suppose there is a buy market order of stock $j$ executed at time $t$. Now, news comes in that triggers sell orders of stock $i$ between the time $t$ and $t+\tau$. To which degree is the price impact of stock $i$ from the trades of stock $j$ weakened due to the news from time $t$ to $t+\tau$?

To answer this question, we recall the definition of the cross--response functions in-- and excluding $\varepsilon_j(t)=0$,
\begin{equation}
R_{ij}^\textrm{(inc. 0)}(\tau)=\frac{\sum_{t=1}^{T_j+T_{j;n}}r_i(t,\tau)\varepsilon_j(t)}{T_j+T_{j;n}} \ ,
\label{eq53}
\end{equation}
\begin{equation}
R_{ij}^\textrm{(exc. 0)}(\tau)=\frac{\sum_{t=1}^{T_j}r_i(t,\tau)\varepsilon_j(t)}{T_j} \ .
\label{eq54}
\end{equation}
For stock $j$, $T_j$ and $T_{j;n}$ are the total trading times of stock $j$ and the total time of lack of trading or of a buy--sell balance, respectively. If trading does not take place or if there is a buy--sell balance, the products $r_i(t,\tau)\varepsilon_j(t)$ vanish.  Thus, the numerators in Eqs.~\eqref{eq53} and~\eqref{eq54} are the same, while the denominators, \textit{i.e}, the normalization constants, differ. Hence, we find
\begin{equation}
R_{ij}^\textrm{(inc. 0)}(\tau)=f_jR_{ij}^\textrm{(exc. 0)}(\tau) \ ,
\label{eq55}
\end{equation}
where we defined the relative trading frequency 
\begin{equation}
f_j=\frac{T_j}{T_j+T_{j;n}} \ .
\label{eq56}
\end{equation}
In parts of the literature, the term frequency is used in a colloquial sense, for example: a frequently traded stock is a very often traded stock.  The definition~\eqref{eq56} is consistent with that, but quantifies it. The most frequently traded stocks have $f_j=1$, because the time $T_{j;n}$ is zero. The longer this time, the lower is the relative trading frequency. Put differently, $f_j$ can be regarded as the probability for trades to occur on the time scale of one second. Coming back to the above question, $f_j$ rescales the degree of impact: the higher the trading frequency $f_j$, the stronger is the impact of trades on the price after a time lag. According to Eq.~\eqref{eq55}, the cross--response including $\varepsilon_j(t)=0$ is the one excluding $\varepsilon_j(t)=0$ scaled by a proper probability. Obviously, the cross--response including $\varepsilon_j(t)=0$ can never be stronger than the one excluding $\varepsilon_j(t)=0$.

We now turn to the average cross--responses. Passive and active cross--responses behave rather differently. For the passive cross--response including $\varepsilon_j(t)=0$, we have
\begin{eqnarray}
R_{i}^{(p,~\textrm{inc. 0})}(\tau) &=& \frac{1}{k}\sum_{j=1}^{k}R_{ij}^\textrm{(inc. 0)}(\tau)
                                                         \nonumber\\
              &=& \frac{1}{k}\sum_{j=1}^{k}f_jR_{ij}^\textrm{(exc. 0)}(\tau) \ ,
\label{eq57}
\end{eqnarray}
where $k$ is the total number of stocks $j$. The cross--respons\-es excluding $\varepsilon_j(t)=0$ weighted with the factors $f_j$ have to be summed over different stocks $j$ to yield the passive cross--response including $\varepsilon_j(t)=0$. Remarkably, the difference of the stock ranks based on the passive cross--responses in-- and excluding $\varepsilon_j(t)=0$ is reduced by this average. On the other hand, for the active cross--response, we find
\begin{eqnarray}
R_{j}^{(a,~\textrm{inc. 0})}(\tau)&=& \frac{1}{k}\sum_{i=1}^{k} R_{ij}^\textrm{(inc. 0)}(\tau)
                                                          \nonumber\\
           &=& \frac{1}{k}\sum_{i=1}^{k}f_jR_{ij}^\textrm{(exc. 0)}(\tau) 
                                                          \nonumber\\
                            &=& f_jR_j^{(a,~\textrm{exc. 0})}(\tau) \ ,
\label{eq58}
\end{eqnarray}
because the factor $f_j$ is independent of the average of the cross--response excluding $\varepsilon_j(t)=0$. The active cross--response including $\varepsilon_j(t)=0$ is simply proportional to the one excluding $\varepsilon_j(t)=0$. This explains that the active cross--response including $\varepsilon_j(t)=0$ exhibits a high correlation with the average daily number of trades while the one excluding $\varepsilon_j(t)=0$ does not.

\section{Comparisons of self-- and cross--responses}
\label{section6}

\begin{figure*}[htbp]
 \begin{center}
    \includegraphics[width=0.8\textwidth]{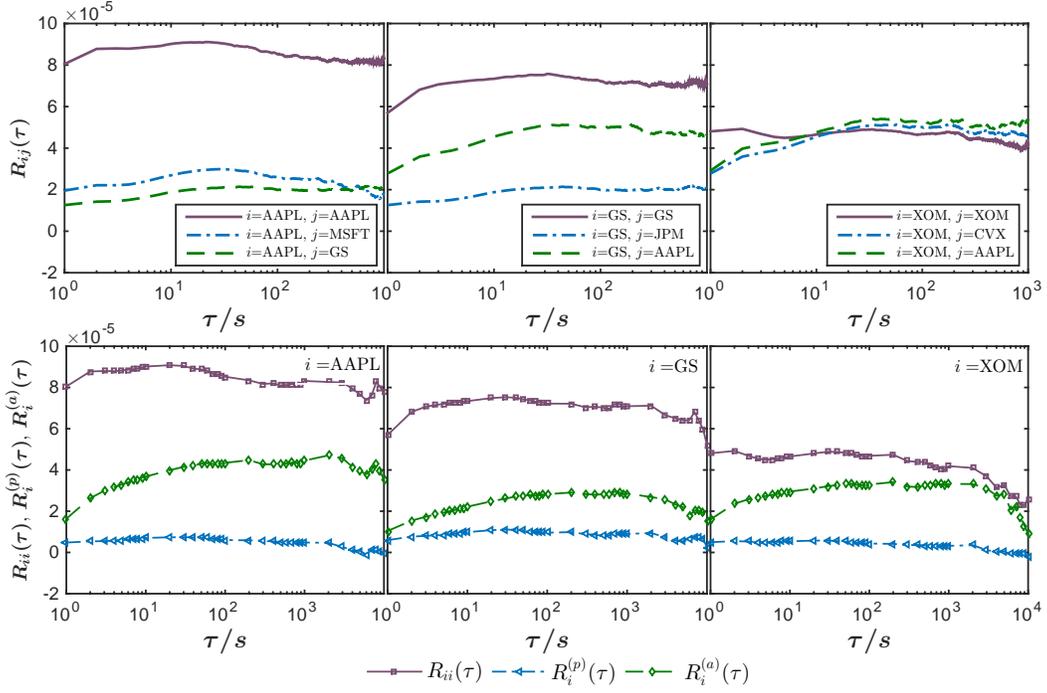}
 \end{center}
 \caption{Comparisons of the self--responses for AAPL, GS and XOM with cross--responses to different stocks (top) versus the time lag $\tau$. Comparisons of self-responses, passive cross--responses, and active cross--responses for the same stocks (bottom) versus the time lag $\tau$.  The scale on the horizontal axes is logarithmic.}
 \label{fig61}
\end{figure*}
\begin{figure*}[htbp]
 \begin{center}
    \includegraphics[width=0.8\textwidth]{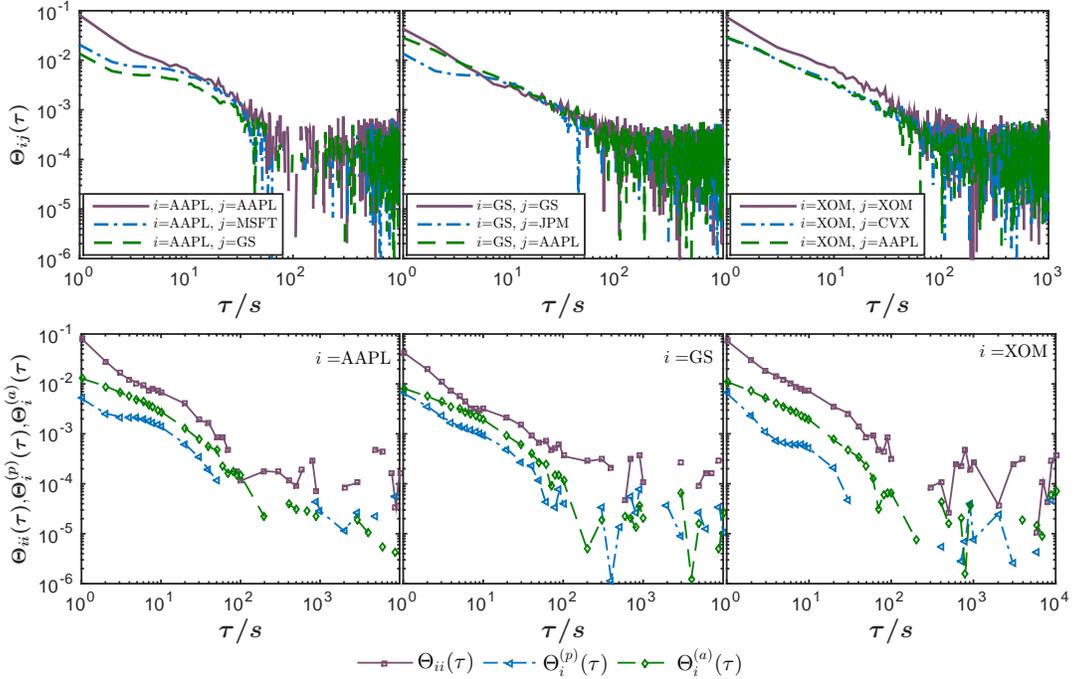}
 \end{center}
 \caption{Comparisons of the trade sign self--correlators for AAPL, GS and XOM with cross--correlators with different stocks (top) versus the time lag $\tau$. Comparisons of the self--correlators, passive cross--correlators, and active cross--correlators for the same stocks (bottom) versus the time lag $\tau$. The scales on all axes are logarithmic. Negative values are suppressed.}
\label{fig62}
\end{figure*}

We compared the self-- and cross--responses in a previous study~\cite{Wang2016}, focussing on the difference of self--responses on different time scales and on the difference of the self-- and cross--responses on the same time scale. Here, we discuss the average cross--responses, and further clarify what happens on the physical time scale. The cross--responses as well as the sign cross--correlators for the stocks of AAPL, GS and XOM are displayed in Figs.~\ref{fig61} and \ref{fig62}. The zero trade signs are included in the analysis, \textit{i.e.}, the influence of trading frequency is taking into account.

Figure~\ref{fig61} shows the comparisons of the self-- and cross--responses as well as the average cross--responses, \textit{i.e.}, passive and active cross--responses. Typically, the self--respon\-se is stronger than the cross--response for two stocks due to the strong self--correlation of trade signs. The example of XOM, however, shows that the cross--response can be stronger than the self--response. The average cross--responses are always weaker than the self--response, implying that cases such as XOM are rare. Due to the noise reduction we can follow the average cross--responses over a longer time interval. On shorter scales of the time lag $\tau$, both, self-- and cross--responses, should be considered when looking at individual stocks, but for investigating cross--response stability, persistence or efficiency of the market as a whole, the average quantities give useful information on longer scales.

In Fig.~\ref{fig62} the trade sign self-- and cross--correlators corresponding to Fig.~\ref{fig61} are depicted. The differences between these correlators are observable for time lags of less than 10 seconds. With the lag increasing, the self--correlators are close to the cross--correlators for AAPL and GS. The exceptional behavior of XOM, however, shows up again in the sign correlators. In the study of the self--response~\cite{Bouchaud2004}, a `bare' impact function with a power--law decay for a single trade was proposed to account for the amplification effect of sign self--correlations. It derives from the self--correlation accumulation with the time lag increasing. Thus, the impact function and the sign self--correlation, mutually describe the price self--response. For XOM, the sign self--correlator is larger than the cross--correlators before the appearance of correlation fluctuations, but its cross--response is stronger than the self--response. This implies that the impact functions of the self-- and cross--responses are different. In other words, there are different impact mechanisms between the self--and cross--responses. 

The averages that we perform produce the combined impacts of different stocks on individual stocks. The bare impact functions cannot be observed directly. Here, we observe the decay of the average sign cross--correlators. The passive and active cross--correlators are always smaller than the self--correlators for AAPL, GS and XOM before the appearance of correlation fluctuations, which is consistent with the case of the cross--responses. Thus, the combined impact function of the individual stocks is more stable than the ones for each stock pair.

\section{Conclusions}  
\label{section7}

We introduced average cross--response functions, a passive and an active one, measuring the average price change of a given stock due to the trades of all others as well as the impact of trading a specific stock on the average price change of the other ones, respectively. Interestingly, the passive cross--response reverses at a relatively short time lag of dozens of seconds or so and then declines rather quickly in a volatile way, while the active cross--response reverses at a longer time lag of some hundreds of seconds with less volatility. This is so, because the price change in one stock easily alerts the market participants. The dispersion over different stocks makes it more difficult to detect an effect due to the noise interference. The average cross--responses considerably reduce this noise, and make generic effects visible. We also introduced the corresponding active and passive trade sign cross--correlators. It is quite remarkable that the short memory for a stock pair turns into a long memory when averaged over different stock pairs.

We compared the cross--responses in-- and excluding the zero trade signs. The cross--response including the zero trade signs reflects the weakened impact of trades due to a lack of trading or due to the balance of buy and sell market orders in one second. We showed that it is the cross--response excluding the zero trade signs, \textit{i.e.} the price impact purely caused by trades, scaled by a trading frequency on the time scale of one second. Which stocks are identified as influencing by studying the active cross--responses differs for in-- and excluding zero trade signs. There is a high correlation with the average daily number of trades when including zero trade signs, but low correlation when excluding zero trade signs.

We also compared the self--response with the various cross--responses. On shorter scales of the time lag, it is useful to consider the self-- and the cross--responses for individual stocks. However, on longer scales, the average cross--responses give interesting new information for investigating the stability, persistence or the efficiency of the market as a whole. Moreover, the comparison of sign self-- and cross--correlators reveals the existence of different impact mechanisms.

\section*{Acknowledgements}

We thank D.~Chetalova, T.A.~Schmitt, Y.~Stepanov, and D.~Wagner for fruitful discussions. We are grateful to the referee of an earlier version  of the paper for several helpful comments. One of us (S.W.) acknowledges financial support from the China Scholarship Council (grant no. 201306890014).

\section*{Author contribution statement}

T.G.~proposed the research. R.S.~and S.W.~developed the method of analysis, which S.W.~carried out. All authors contributed equally to analyzing the results, the paper was written by S.W.~and T.G.

\appendix

\section{Stocks used for analyzing the market response}
\label{appA}

The market response structure and the identifying of influencing and influenced stocks are based on following 99 stocks from ten economic sectors: industrials (I), health care (HC), consumer discretionary (CD), information technology (IT), utilities (U), financials (F), materials (M), energy (E), consumer staples (CS), and  telecommunications services (TS) as listed in Table~\ref{tabA}. The acronym AMC in Table~\ref{tabA} stands for averaged market capitalization.
\begin{table*}
\linespread{0.5}
\caption{Information of 99 stocks from ten economic sectors} 
\begin{center}
\begin{footnotesize}
\begin{tabular}{llrc@{\hskip 0.4in}llr} 
\hline
\hline
\\
\multicolumn{3}{l}{Industrials (I)} &~~& \multicolumn{3}{l}{Financials (F)}   \\
\cline{1-3}\cline{5-7}
Symbol	&Company				&AMC~ 	&	&Symbol		&Company					&AMC~	\\
\cline{1-3}\cline{5-7}
FLR		&Fluor Corp. (New)			&14414.4	&	&CME		&CME Group Inc.				&49222.9	\\
LMT		&Lockheed Martin Corp.		&12857.8	&	&GS			&Goldman Sachs Group			&21524.3	\\
FLS		&Flowserve Corporation		&12670.2	&	&ICE			&Intercontinental Exchange Inc.	&14615.3	\\
PCP		&Precision Castparts			&12447.0	&	&AVB		&AvalonBay Communities			&11081.6	\\
LLL		&L-3 Communications Holdings&12170.8	&	&BEN		&Franklin Resources				&10966.2	\\
UNP		&Union Pacific				&11920.9	&	&BXP		&Boston Properties				&10893.0 	\\
BNI		&Burlington Northern Santa Fe C &11837.5&	&SPG		&Simon Property 	Group  Inc	 	&10862.4	\\
FDX		&FedEx Corporation			&10574.7	&	&VNO		&Vornado Realty Trust			&10802.3	\\
GWW	&Grainger (W.W.) Inc.		&10416.8	&	&PSA		&Public Storage				&10147.9	\\
GD		&General Dynamics			&10035.6	&	&MTB		&M$\&$T Bank Corp.			&9920.2	\\
\cline{1-3}\cline{5-7}
\\
 \multicolumn{3}{l}{Health Care (HC)} &~& \multicolumn{3}{l}{Materials (M)}   \\
\cline{1-3}\cline{5-7}
Symbol	&Company				&AMC~ 	&	&Symbol		&Company					&AMC~	\\
\cline{1-3}\cline{5-7}
ISRG	&Intuitive Surgical Inc.		&31355.9	&	 &X			&United States Steel Corp.		&15937.7	\\
BCR		&Bard (C.R.) Inc.			&11362.7	&	 &MON		&Monsanto Co.					&14662.6	\\
BDX		&Becton  Dickinson			&10298.4	&	 &CF			&CF Industries Holdings Inc		&14075.5	\\
GENZ	&Genzyme Corp.			&9728.8	&	 &FCX		&Freeport-McMoran Cp $\&$ Gld 	&11735.7	\\
JNJ		&Johnson $\&$ Johnson		&9682.6	&	&APD		&Air Products $\&$ Chemicals		&10246.4	\\
LH		&Laboratory Corp. of America Holding&9035.7& &PX		&Praxair  Inc.					&10234.5	\\
ESRX	&Express Scripts			&8864.6	&	&VMC		&Vulcan Materials				&8700.4	\\
CELG	&Celgene Corp.			&8783.1	&	&ROH		&Rohm $\&$ Haas				&8527.1	\\
ZMH		&Zimmer Holdings			&8681.7	&	&NUE		&Nucor Corp.					&7997.4	\\
AMGN	&Amgen					&8543.0	&	&PPG		&PPG Industries				&7336.7	\\
\cline{1-3}\cline{5-7}
\\
\multicolumn{3}{l}{Consumer	Discretionary (CD)} &~& \multicolumn{3}{l}{Energy (E)}   \\
\cline{1-3}\cline{5-7}
Symbol	&Company				&AMC~ 	&	&Symbol		&Company					&AMC~	\\
\cline{1-3}\cline{5-7}
WPO	&Washington Post			&61856.1	&	&RIG		&Transocean Inc. (New)			&16409.5	\\
AZO		&AutoZone Inc.				&14463.7	&	&APA		&Apache Corp.					&13981.9	\\
SHLD	&Sears Holdings Corporation	&11759.2	&	&EOG		&EOG Resources				&13095.0	\\
WYNN	&Wynn Resorts Ltd.			&11507.9	&	&DVN		&Devon Energy Corp.			&12499.7	\\
AMZN	&Amazon Corp.			&10939.2	&	&HES		&Hess Corporation				&11990.4	\\
WHR	&Whirlpool Corp.			&9501.9	&	&XOM		&Exxon Mobil Corp.				&11460.3	\\
VFC		&V.F. Corp.				&9051.2	&	&SLB		&Schlumberger Ltd.				&11241.1	\\
APOL	&Apollo Group				&8495.8	&	&CVX		&Chevron Corp.				&11100.0	\\
NKE		&NIKE Inc.				&8149.5	&	&COP		&ConocoPhillips				&10215.3	\\
MCD		&McDonald's Corp.			&8025.6	&	&OXY		&Occidental Petroleum			&9758.4	\\
\cline{1-3}\cline{5-7}
\\
\multicolumn{3}{l}{Information Technology (IT)} &~& \multicolumn{3}{l}{Consumer Staples (CS)}   \\
\cline{1-3}\cline{5-7}
Symbol	&Company				&AMC~ 	&	&Symbol		&Company					&AMC~	\\
\cline{1-3}\cline{5-7}
GOOG	&Google Inc.				&62971.6	&	&BUD		&Anheuser-Busch				&9780.6	\\
MA		&Mastercard Inc.			&28287.8	&	&PG			&Procter $\&$ Gamble			&9711.5	\\
AAPL	&Apple Inc.				&22104.1	&	&CL			&Colgate-Palmolive				&9549.2	\\
IBM		&International Bus. Machines	&15424.9	&	&COST		&Costco Co.					&9545.9	\\
MSFT	&Microsoft Corp.			&10845.1	& 	&WMT		&Wal-Mart Stores				&9325.7	\\
CSCO	&Cisco Systems			&8731.4	&	&PEP		&PepsiCo Inc.					&9180.7	\\
INTC		&Intel Corp.				&8385.8	&	&LO			&Lorillard Inc.					&8919.0	\\
QCOM	&QUALCOMM Inc.			&7739.4	&	&UST		&UST Inc.						&8433.1	\\
CRM		&Salesforce Com Inc. 		&7691.9	&	&GIS		&General Mills					&8243.3	\\
WFR		&MEMC Electronic Materials	&7392.8	&	&KMB		&Kimberly-Clark				&8069.5	\\
\cline{1-3}\cline{5-7}
\\
\multicolumn{3}{l}{Utilities (U)} &~& \multicolumn{3}{l}{Telecommunications Services (TS)}   \\
\cline{1-3}\cline{5-7}
Symbol	&Company				&AMC~ 	&	&Symbol		&Company					&AMC~	\\
\cline{1-3}\cline{5-7}
ETR		&Entergy Corp.				&12798.7	&	&T			&AT$\&$T Inc.					&6336.2	\\
EXC		&Exelon Corp.				&9738.8	&	&VZ			&Verizon Communications		&5732.5	\\
CEG		&Constellation Energy  Group 	&9061.5	&	&EQ			&Embarq Corporation			&5318.7	\\
FE		&FirstEnergy Corp.			&8689.4	&	&AMT		&American Tower Corp.			&5195.6	\\
FPL		&FPL Group				&7742.8	&	&CTL 		&Century Telephone				&4333.8	\\
SRE		&Sempra Energy			&6940.6	&	&S			&Sprint Nextel Corp.				&2533.7	\\
STR		&Questar Corp.				&6520.4	&	&Q			&Qwest Communications  Int 		&2201.3	\\
TEG		&Integrys Energy Group Inc. 	&5978.4	&	&WIN		&Windstream Corporation			&2089.1	\\
EIX		&Edison Int'l				&5877.5	&	&FTR		&Frontier Communications		&1580.9	\\
AYE		&Allegheny Energy			&5864.9	&	&			&							&		\\
\hline
\hline
\end{tabular}
\end{footnotesize}
\end{center}
\label{tabA}
\end{table*}

\end{document}